\documentclass[twocolumn,printnumbers,amsmath,amssymb,prl]{revtex4}
\usepackage{graphicx}
\usepackage{color}

\begin{document}
\title{Two-scale scenario of rigidity percolation of sticky particles}

\author{Yuchuan Wang$^{1,2}$}
\author{Sheng Fang$^{1,3}$}
\author{Ning Xu$^{1,2,*}$}
\author{Youjin Deng$^{1,3,\dagger}$}

\affiliation{$^1$Hefei National Laboratory for Physical Sciences at the Microscale,  Hefei 230026, P. R. China; \\$^2$CAS Key Laboratory of Microscale Magnetic Resonance and Department of Physics, University of Science and Technology of China, Hefei 230026, P. R. China; \\$^3$CAS Center for Excellence and Synergetic Innovation Center in Quantum Information and Quantum Physics and Department of Modern Physics, University of Science and Technology of China, Hefei 230026, P. R. China}

\begin{abstract}
In the presence of attraction, the jamming transition of packings of frictionless particles corresponds to the rigidity percolation. When the range of attraction is long, the distribution of the size of rigid clusters, $P(s)$, is continuous and shows a power-law decay. For systems with short-range attractions, however, $P(s)$ appears discontinuous. There is a power-law decay for small cluster sizes, followed by a low probability gap and a peak near the system size. We find that this appearing ``discontinuity'' does not mean that the transition is discontinuous. In fact, it signifies the coexistence of two distinct length scales, associated with the largest cluster and smaller ones, respectively. The comparison between the largest and second largest clusters indicates that their growth rates with system size are rather different. However, both cluster sizes tend to diverge in the large system size limit, suggesting that the jamming transition of systems with short-range attractions is still continuous. In the framework of the two-scale scenario, we also derive a generalized hyperscaling relation. With robust evidence, our work challenges the former single-scale view of the rigidity percolation.
\end{abstract}

\maketitle

Particulate systems, e.g., granular materials and colloids, turn into disordered solids via the jamming transition upon compression \cite{liu1}. In the past two decades, people have made significant progress to understand the jamming transition, by studying packings of frictionless spheres interacting via repulsions \cite{ohern,torquato1,liu2,hecke,torquato2,parisi,xu1,xu2,wyart,muller,charbonneau,durian,olsson,silbert,donev}. Recently, the jamming of sticky particles has attracted much attention, which exhibits different behaviors from repulsive particles \cite{trappe,lois,zheng,koeze,lu1,lu2,irani}. Compared with repulsive systems which are somehow idealized, the study of attractive systems has not only theoretical but also practical merits, because attraction is often present in real systems.

Percolation is one of the particular issues of the jamming transition \cite{lois,koeze,pathak,yang,tong}. It has been shown that upon compression a packing of sticky particles undergoes connectivity and rigidity percolation transitions, which belong to new percolation universality classes \cite{lois}. Recently, the jamming transition of sticky particles has been investigated from the perspective of rigidity percolation \cite{koeze}. It has been found that for finite-size systems the distribution of the size of rigid clusters exhibits a discontinuity when attraction is weak, while for strong attractions it is continuous.  From finite size scaling of the mean size of nonspanning clusters, it has been proposed that sufficiently large systems with any strength of attraction will fall in the strongly attractive universality class. Therefore, no matter how weak the attraction is, the jamming transition of sticky particles is continuous.

In this Letter, we revisit the rigidity percolation of sticky particles, with special attention to the ``discontinuous'' distribution of the size of rigid clusters, $P(s)$. When the range of attraction is short, there is a low probability gap in $P(s)$ between the largest cluster and smaller ones \cite{koeze}. The gap is more pronounced when the range of attraction decreases. The distribution of the largest cluster size is peaked at a size $s_{\rm m}$ near the system size with a peak value $P_{\rm max}$, while for smaller clusters $P(s)\sim s^{-\tau}$. We find that $P_{\rm max}\sim s_{\rm m}^{-\tau_{\rm m}}$ with the variation of system size. For systems with relatively long-range attractions and a continuous $P(s)$, $\tau=\tau_{\rm m}$. For systems with short-range attractions and a seemingly ``discontinuous'' $P(s)$, $\tau>\tau_{\rm m}$. When the range of attraction decreases, $\tau$ becomes larger. Focusing on the largest and second largest clusters, we find that they both grow when system size increases, but showing different scalings. For both clusters, our analysis shows that the critical exponents satisfy a generalized hyperscaling relation. Therefore, the appearing ``discontinuity'' in $P(s)$ does not imply that the transition is discontinuous. With short-range attractions, the rigidity percolation transition is continuous, but governed by two distinct length scales.

Our systems are essentially the same as in previous work \cite{lois,zheng,koeze}. We consider two-dimensional systems with a side length $L$ and periodic boundary conditions in both directions. To avoid crystallization, we adopt a binary mixture of $N/2$ large and $N/2$ small disks with a diameter ratio of $1.4$. The interparticle potential is
\begin{equation}
U(r_{ij})=
\begin{cases}
\frac{\epsilon}{2}\left[\left( 1-\frac{r_{ij}}{\sigma_{ij}}\right)^{2}-2\mu^{2}\right],\frac{r_{ij}}{\sigma_{ij}}\le 1+\mu,\\
-\frac{\epsilon}{2}\left( 1+2\mu-\frac{r_{ij}}{\sigma_{ij}}\right)^{2},\ \ 1+\mu<\frac{r_{ij}}{\sigma_{ij}}\le 1+2\mu,\\
0,\ \ \ \ \ \ \ \ \ \ \ \ \ \ \ \ \ \ \ \ \ \ \ \ \ \ \ \frac{r_{ij}}{\sigma_{ij}}>1+2\mu,\label{potential}
\end{cases}
\end{equation}
where $r_{ij}$ and $\sigma_{ij}$ are the separation between particles $i$ and $j$ and sum of their radii, and $\mu$ controls the range and strength of attraction \cite{note_range}. The interaction becomes purely repulsive when $\mu=0$. We set the unit of length to be the small particle diameter $\sigma$.

\begin{figure}
  \includegraphics[width=\columnwidth]{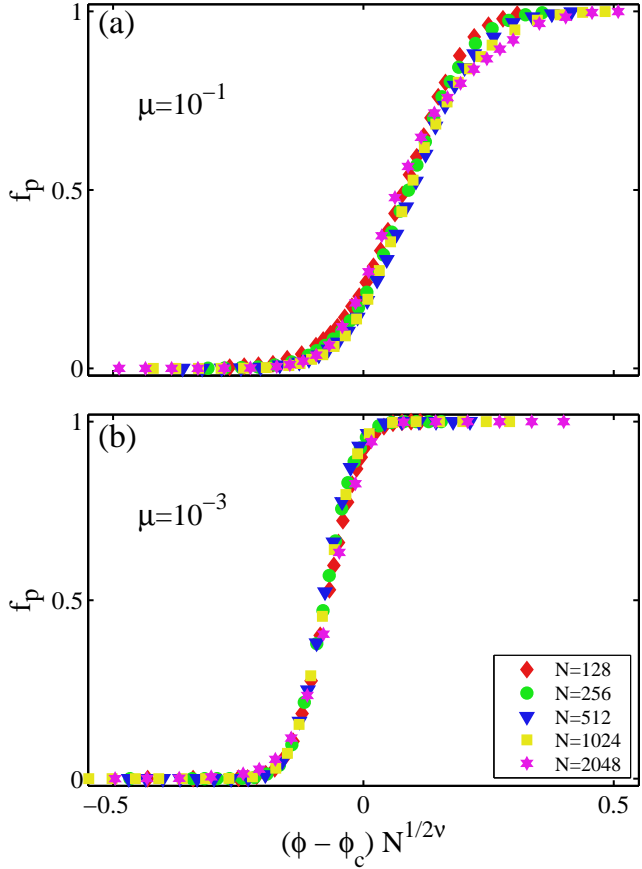}
  \caption{\label{fig:fig1}Finite size scaling of the probability of finding jammed states with a percolating rigid cluster, $f_{\rm p}$, as a function of packing fraction $\phi$. For $\mu=10^{-1}$ and $10^{-3}$,
  the scaling exponents $\nu$ used here are $2.30$ and $1.08$, respectively. $\phi_{\rm c}$ is the critical packing fraction of the jamming transition in the thermodynamic limit.  }
  \label{fig:distortion}
\end{figure}

At fixed packing fraction $\phi$, we quickly quench high-temperature configurations to local potential energy minima using the fast inertial relaxation engine method \cite{fire}. For each minimum, we identify rigid clusters using the pebble game algorithm \cite{pebble1,pebble2,pebble3}. The minimum is rigid (jammed) if there is a percolating rigid cluster. Figure~\ref{fig:fig1} shows finite size scaling of the probability of finding rigid states, $f_{\rm p}(\phi, N)$, for $\mu=10^{-1}$ and $10^{-3}$. Consistent with previous studies \cite{koeze}, curves of different $N$ collapse when $f_{\rm p}$ is plotted against $(\phi - \phi_{\rm c}) N^{1/2\nu}$. Here $\phi_{\rm c}$ is the critical packing fraction of the jamming transition in the thermodynamic limit, and $\nu$ is estimated from $\delta \phi \sim N^{-1/2\nu}$ with $\delta \phi$ being the width of $f_{\rm p}(\phi,N)$.

Figure~\ref{fig:fig1} suggests that the jamming transition is associated with the rigidity percolation transition for sticky particles. However, the values of $\nu$ are quite different for the two attractions ($\nu=2.30\pm0.20$ and $1.08\pm0.18$ for $\mu=10^{-1}$ and $10^{-3}$, respectively). This already suggests that the rigidity percolation at $\mu=10^{-1}$ and $10^{-3}$ belong to {\it different} universality classes.

\begin{figure}
  \includegraphics[width=\columnwidth]{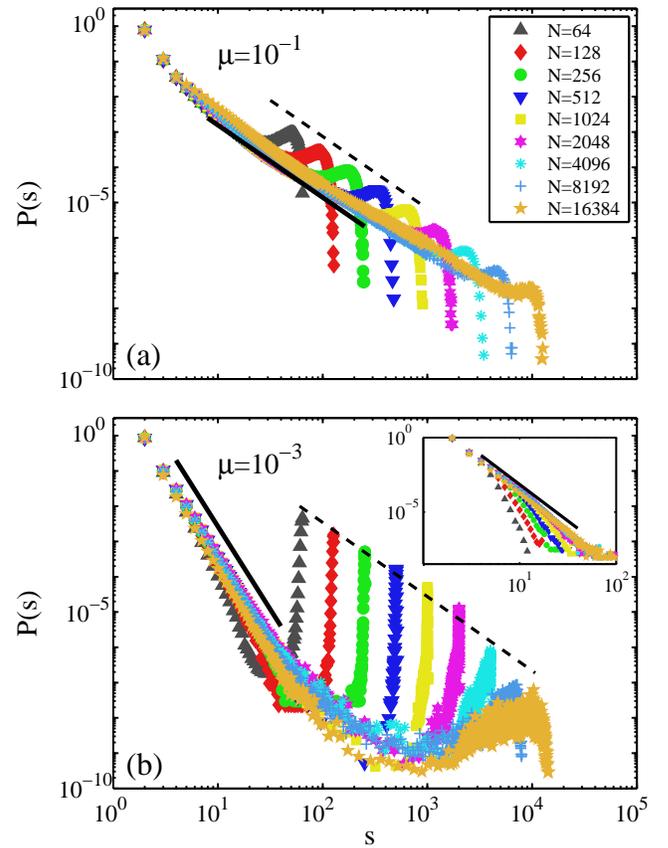}
  \caption{\label{fig:fig2} System size evolution of the distribution of the size of rigid clusters, $P(s)$, calculated at $f_{\rm p}\approx 0.5$. The inset of (b) shows the distributions with the largest cluster being removed. The solid and dashed lines have a slope of $-\tau$ and $-\tau_{\rm m}$, respectively. In (a), we plot the lines with $\tau=\tau_{\rm m}= 2.05$. In (b), $\tau=4.7$ and $\tau_{\rm m}= 2.1$. }
  \label{fig:distortion}
\end{figure}

Figure~\ref{fig:fig2} shows the distribution of the size of rigid clusters, $P(s)$, at $f_{\rm p}\approx 0.5$ for $N$ ranging from $64$ to $16384$. When $\mu=10^{-1}$, Fig.~\ref{fig:fig2}(a) shows that $P(s)\sim s^{-\tau}$, with $\tau= 2.05\pm0.05$, in agreement with previous studies \cite{lois,koeze}. Note that there is a peak at a large value of $s$, but the whole distribution curve can be treated as continuous. When $\mu=10^{-3}$, as shown in Fig.~\ref{fig:fig2}(b), $P(s)$ splits into two relatively discrete parts. For small clusters, $P(s)\sim s^{-\tau}$, with $\tau$ being apparently larger than that of $\mu=10^{-1}$. For large values of $s$ close to $N$, $P(s)$ exhibits a sharp peak at $s_{\rm m}$, contributed by the largest cluster of all states. In the Supplemental Material (SM) \cite{sm}, we also show $P(s)$ for $\mu=10^{-5}$, which looks more discrete and has a larger $\tau$ than $\mu=10^{-3}$.

The appearing ``discontinuity'' in $P(s)$ violates our conventional perception of rigidity percolation \cite{pebble1,lois}, probably leading to an argument that it signifies a discontinuous transition. Moreover, when $\mu=10^{-3}$ and $10^{-5}$, the values of $\tau$ are $4.7\pm0.2$ and $5.6\pm0.2$, respectively. Assuming hyperscaling relation, $D(\tau -1)=d$, still holds, the fractal dimension $D$ would be $0.54\pm0.03$ and $0.43\pm0.02$, respectively, where $d=2$ is the dimension of space.  These $D$ values are too small to be reasonable fractal dimensions. Therefore, a possible explanation is that the hyperscaling relation is not satisfied \cite{lois}, which may further hint that the transition is discontinuous. However, we will evidence that this ``discontinuity'' does not violate the continuous nature of rigidity percolation. With correct characterization, it actually reveals a new percolation scenario with two length scales.

The inset of Fig.~\ref{fig:fig2}(b) shows $P(s)$ at $\mu=10^{-3}$ with the largest cluster being excluded. When $N$ increases, $\tau$ shows no trend of decrease. Therefore, with current computational power, we believe that the ``abnormally'' large values of $\tau$ for systems with short-range attractions are not finite size effect.

Another noticeable feature of $P(s)$ at $\mu=10^{-3}$ is that the peak at $s_{\rm m}$ decays when $N$ increases. If we denote the peak value as $P_{\rm max}$, Fig.~\ref{fig:fig2}(b) shows that $P_{\rm max}\sim s_{\rm m}^{-\tau_{\rm m}}$.  Interestingly, $\tau_{\rm m}=2.1\pm0.1$, close to the ``normal'' $\tau$ value as for $\mu=10^{-1}$, but much smaller than $\tau=4.7\pm0.2$. Note that for $\mu=10^{-1}$ Fig.~\ref{fig:fig2}(a) also shows that $P_{\rm max}\sim s_{\rm m}^{-\tau_{\rm m}}$ with $\tau_{\rm m}\approx \tau$. Therefore, the coexistence of $\tau$ and $\tau_{\rm m}$ is general.

\begin{figure}
  \includegraphics[width=\columnwidth]{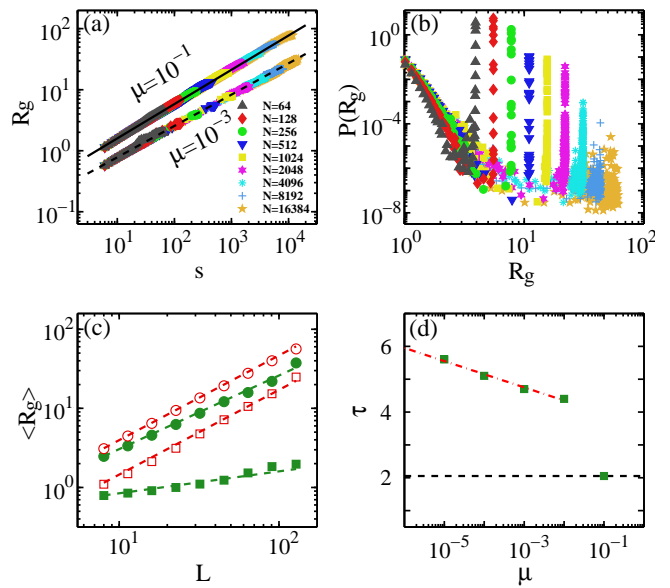}
  \caption{\label{fig:fig3}  (a) Gyration radius $R_{\rm g}$ versus cluster size $s$. The top and bottom branches are for $\mu=10^{-1}$ and $10^{-3}$, respectively. To distinguish the two branches, we show $R_{\rm g}/2$ for $\mu=10^{-3}$. The solid and dashed lines have a slope of $1/d_{\rm f}$ with $d_{\rm f}=1.781$ and $1.953$, respectively. (b) System size dependence of the distribution of the gyration radius $P(R_{\rm g})$. The legend in (a) applies to both (a) and (b). (c) Average gyration radius $\langle R_{\rm g}\rangle$ of the largest (circles) and second largest (squares) clusters versus the side length of the system $L\sim N^{1/2}$. The empty and solid symbols are for $\mu=10^{-1}$ and $10^{-3}$, respectively. The dashed lines show the predicted scaling relation $R_{\rm g}\sim L^{\kappa}$, as discussed in the text, with $\kappa$ being calculated from the generalized hyperscaling relation using values of $\tau_{\rm m}$, $\tau$, and $d_{\rm f}$. The dashed lines shown here are with $\kappa_1=1.07$ and $\kappa_2=1.07$ for the largest and second largest clusters when $\mu=10^{-1}$, and $\kappa_1=0.93$ and $\kappa_2=0.28$ when $\mu=10^{-3}$. (d) Dependence of $\tau$ on the range of attraction $\mu$. The horizontal dashed line is $\tau=2.05$, showing the one-scale limit. The dot-dashed line is a logarithmic fit to $\tau(\mu)$ for small values of $\mu$.}
  \label{fig:distortion}
\end{figure}

To characterize self-similar geometric clusters, the fractal dimension $d_{\rm f}$ is defined to describe how the size $s$ of a cluster grows with its linear size $\xi$ as $s \sim \xi^{d_{\rm f}}$. Here, we distinguish $d_{\rm f}$ from $D$ for the reasons to be discussed later, and calculate the gyration radius $R_{\rm g}$ for each cluster to represent $\xi$. Figure~\ref{fig:fig3}(a) shows $R_{\rm g}$ versus $s$ for all clusters in systems with different $N$, which exhibit two interesting features.

{\it First}, all clusters follow a universal scaling law $R_{\rm g} \sim s^{1/d_{\rm f}}$. This suggests that given any range of attraction the geometric structures of all critical clusters are self-similar and can be described by a single fractal dimension $d_{\rm f}$. Taken into account how the largest cluster behaves differently from the others in $P(s)$, the self-similiarity of all clusters are rather surprising, stimulating us to examine the physical meaning of the ``fractal dimension" $D=d/(\tau - 1)$.

{\it Second}, the fractal dimension $d_{\rm f}$ weakly depends on the range of attraction. Seen from Fig.~\ref{fig:fig3}(a), $d_{\rm f}= 1.781\pm0.003$ and $1.953\pm0.007$ for $\mu=10^{-1}$ and $10^{-3}$, respectively. This difference again suggests that the rigidity percolation transitions at $\mu=10^{-1}$ and $10^{-3}$ belong to {\it different} universality classes.

In the study of critical phenomena, the standard scaling theory assumes that near criticality there exists {\it only one} characteristic length $\xi$, diverging as $\xi \sim |t|^{1/\nu}$ with $t$ being the distance to the critical point. Further, it is assumed that, for a finite system with a linear size $L\sim N^{1/d}$, $\xi\approx L$, so that the size of the largest cluster grows as $\xi^{d_{\rm f}} \approx L^{d_{\rm f}}$.
However, a careful examination of Fig.~\ref{fig:fig3}(a) shows that for $\mu=10^{-3}$ a gap gradually occurs between $R_{\rm g}$ for the largest cluster and for the others when $N$ increases.
We plot the $R_{\rm g}$ distribution in Fig.~\ref{fig:fig3}(b), which behaves as $P(R_{\rm g})\sim s^{1-1/d_{\rm f}}P(s)$ from $s\sim R_{\rm g}^{d_{\rm f}}$.
As in $P(s)$, a discontinuity is developed for large $N$, giving a direct illustration of more than one length scales.

Figure~\ref{fig:fig3}(c) shows the average gyration radius $\langle R_{\rm g}\rangle$ against $L$ for the largest and second largest rigid clusters. Both radii exhibit a power-law growth with $L$ as $\langle R_{\rm g}\rangle \sim L^{\kappa}$. For $\mu=10^{-1}$, both radii grow approximately linearly with $L$ ($\kappa_1 \approx \kappa_2 \approx 1$), supporting the assumption of a {\it single} characteristic length scale in the standard scaling theory. In contrast, for $\mu=10^{-3}$, while the radius of the largest cluster, $R_{\rm g,1}$, still grows approximately linearly with $L$ ($\kappa_1 \approx 1$), the growth rate of the radius of the second largest cluster, $R_{\rm g,2}$, is much slower with $\kappa_2$ being significantly smaller than $1$. Therefore, despite a single actual fractal dimension $d_{\rm f}$, the average size of the largest cluster, $\left< s_1 \right> \sim \langle R_{\rm g,1}\rangle^{d_{\rm f}} \sim L^{\kappa_1 d_{\rm f}} \approx L^{d_{\rm f}}$,  grows in a much faster rate than that of the second largest one, which has an exponent $\kappa_2 d_{\rm f} < d_{\rm f}$.

With the two length scale scenario, the ``discontinuous" behavior of $P(s)$ for systems with short-range attractions is now a natural consequence, since the growth rates of the largest cluster and the others are different. Moreover, a {\it generalized hyperscaling relation} can be derived via the standard procedure~\cite{percolation_book1,percolation_book2}, and be used to quantitatively account for the $P(s)$ behaviors in Fig.~\ref{fig:fig2}. Given a critical distribution $P(s) \sim s^{-\tau} f(s/s_{\rm c})$ which has a cut-off size $s_{\rm c} \sim \xi_{\rm c}^{d_{\rm f}}$ due to finite systems, it is assumed that there only exist a few number of clusters of sizes near $s_{\rm c}$ with a width $\delta s \sim s_{\rm c}$. Namely, one has  $P(s_{\rm c}) \delta s  \sim \xi_{\rm c}^{-d_{\rm f} (\tau-1)} \sim {\rm O}(L^{-d})$. Making use of $\xi_{\rm c} \sim L^{\kappa}$, we obtain a generalized hyperscaling relation
\begin{equation}
\kappa d_{\rm f} (\tau-1)= d, \label{hyper}
\end{equation}
so that $D=\kappa d_{\rm f}$. Using values of $\tau_{\rm m}$, $\tau$, and $d_{\rm f}$, we are able to obtain expected values of $\kappa_1$ and $\kappa_2$. Equation~(\ref{hyper}) is valid if the expected scalings fit the $\left<R_{\rm g}(L)\right>$ data well. For $\mu=10^{-1}$ with $\tau \approx \tau_m\approx 2.05$ and $d_{\rm f}\approx 1.781$, Eq.~(\ref{hyper}) leads to $\kappa_1\approx \kappa_2\approx 1.07$, so the conventional hyperscaling relation $D(\tau - 1)=d_{\rm f} (\tau-1)= d$ is recovered. For $\mu=10^{-3}$ with $\tau_{\rm m}\approx 2.1$, $\tau\approx 4.7$, and $d_{\rm f}\approx 1.953$, we have $\kappa_1\approx 0.93$ and $\kappa_2\approx 0.28$. As shown by the dashed lines in Fig.~\ref{fig:fig3}(c), the expected values of $\kappa_1$ and $\kappa_2$ can describe the behaviors of $\left< R_{\rm g}(L)\right>$ nicely.

When $\mu=10^{-3}$, the power-law behavior, $\left< R_{\rm g, 2}\right> \sim L^{0.28}$, indicates that the size of the second largest cluster still diverges in the thermodynamic limit. This is a direct evidence suggesting that the jamming transition of systems with short-range attractions is continuous. Nevertheless, different from the previous single length scale picture of percolation, there are two distinct length scales, $\xi_1 \sim L^{\kappa_1} =  L$ for the largest cluster and $\xi_2 \sim L^{\kappa_2} \approx L^{0.28}$ for the others.

Figure~\ref{fig:fig3}(d) shows the evolution of $\tau$ with $\mu$. When $\mu$ decreases, $\tau$ increases, indicating that $\kappa_2$ decreases. It would be expected from Fig.~\ref{fig:fig3}(d) that $\tau\rightarrow \infty$ ($\kappa_2\rightarrow 0$) in the $\mu\rightarrow 0$ limit. If then, as long as $\mu>0$, $\tau$ remains finite and the shorter length diverges in the thermodynamic limit, so that the jamming transition is still continuous. On the other hand, Fig.~\ref{fig:fig3}(d) shows that $\tau$ undergoes a fast change from $\mu=10^{-1}$ to $10^{-2}$, where the percolation evolves from one-scale to two-scale \cite{note_2}.

With special concerns of the ``discontinuous'' feature of the rigid cluster size distribution $P(s)$ for systems with short-range attraction at the rigidity percolation transition, we find that this ``discontinuity'' is not finite size effect and reveals the underlying physics, which challenges the previous understanding of rigidity percolation. The appearing ``discontinuity'' of $P(s)$ is not a signature of discontinuous transition. On the contrary, it signifies the coexistence of two length scales, both diverging in the thermodynamic limit and maintaining the continuity of the transition. Consistent with the previous work \cite{koeze}, our study supports that as long as attraction is present the jamming transition is continuous.

The two-scale scenario is not unique to the jamming transition studied in this work. In a parallel study of rigidity percolation of sticky particles under shear, we find the same scenario below the jamming transition, with the value of $\tau$ increasing when packing fraction increases \cite{wang_next}. Moreover, we notice that two-scale pictures occur frequently in the fields of equilibrium statistical mechanics and condensed matter physics. For the so-called $2$-state random-cluster model on finite complete graph with size $N$ (the Fortuin-Kasterleyn representation of the Ising model), it has been rigorously proved~\cite{Luczak} that at the critical point, while the size $s_1$ of the largest cluster scales as $s_1 \!  \sim \! N^{3/4}$, the size distribution of the second largest cluster follows a function $\tilde{P}(x)$ with $x \equiv s_2 /( \sqrt{N} \log N)$. Similarly, for the critical $2$-state random-cluster model on periodic five-dimensional hypercubes with a linear size $L$, it is numerically shown~\cite{Fang2019}  that the fractal dimension $d_{\rm f}$ for the largest cluster is $3d/4=15/4$ but $d_{\rm f}$ for the others is $1+d/2=7/2$. In the study of quantum deconfined criticality, a phase transition theory beyond the conventional Ginzburg-Landau mechanism, the picture of two diverging length scales, $\xi_1 \! \sim L$ and $\xi_2 \! \sim L^\kappa$ with $0 \! < \! \kappa\! < \! 1$, was introduced~\cite{Shao2016} to account for the Monte Carlo data for a model Hamiltonian for quantum magnet.

Exploration of the underlying mechanisms for the emergence of two length scales is challenging. For quantum deconfined criticality,  it is attributed to the competition of two spontaneous order parameters that possess different symmetries~\cite{Shao2016}. For the complete-graph $2$-state random-cluster model, by introducing an additional parameter such that two distinct fixed points can occur, Ref.~\cite{Fang2020} provides an intuitive explanation from renormalization-flow perspective, and numerically determine the two diverging length/size scales that are respectively associated with each fixed point. One might also consider to study the rigidity percolation of the jamming transition in a broader parameter space. A possibility is to include thermal effects, such that the athermal rigidity percolation becomes unstable and small thermal fluctuations would drive the transition toward some new universality. We wish to leave it for future studies.

We thank Brian Tighe for instructive discussions. This work was supported by the National Natural Science Foundation of China (under Grants No.~11734014 and No.~11625522) and by the National Key R\&D Program of China (under Grants No.~2016YFA0301604 and No.~2018YFA0306501). We also thank the Supercomputing Center of University of Science and Technology of China for the computer time.

\makeatletter
\def\fnum@figure#1{FIG.~S\thefigure$:$~}
\makeatother

\section{Supplemental Material}

\begin{figure}[h]
  \includegraphics[width=0.42\textwidth]{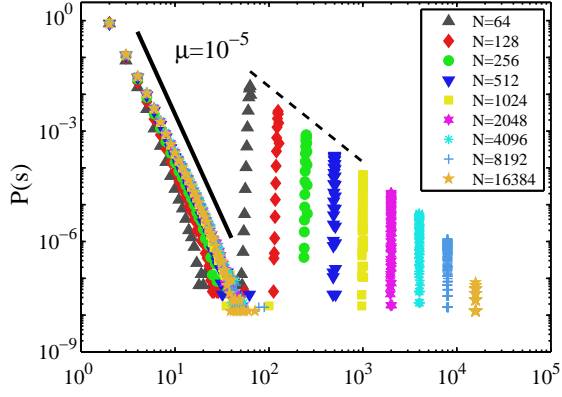}
  \caption{\label{fig:figs1} System size evolution of the distribution of the size of rigid clusters, $P(s)$, for $\mu=10^{-5}$. The solid and dashed lines have a slope of $-\tau$ and $-\tau_m$ with $\tau=5.6$ and $\tau_m= 2.1$, respectively. }
  \label{fig:distortion}
\end{figure}

Figure~S\ref{fig:figs1} shows the distributions of the size of rigid clusters, $P(s)$, for $\mu=10^{-5}$. Because the range of attraction is much shorter than those discussed in the main text, $P(s)$ appears even more ``discontinuous". When cluster size $s$ is small, there is still a power-law decay of $P(s)$: $P(s)\sim s^{-\tau}$ with $\tau=5.6\pm0.2$, which is larger than that for $\mu=10^{-3}$.

\begin{figure}[h]
  \includegraphics[width=0.42\textwidth]{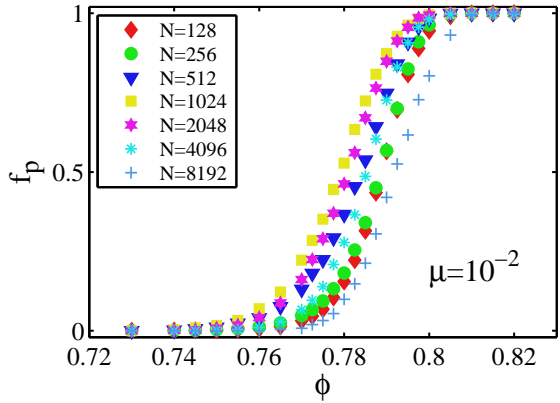}
  \caption{\label{fig:figs2} System size evolution of the probability of rigidity percolation, $f_{\rm p}(\phi)$, for $\mu=10^{-2}$. }
  \label{fig:distortion}
\end{figure}

Figure~S\ref{fig:figs2} suggests that the probability of rigidity percolation $f_{\rm p}(\phi, N)$ for $\mu=10^{-2}$ cannot be collapsed well for system sizes studied in this work. When $N$ increases, the $f_{\rm p}(\phi)$ curves move non-monotonically along the $\phi$ axis, probably due to system size effects. Even worse, the curves do not show the apparent tendency to become steeper with the increase of $N$ and there is no well-defined intersection, which are unexpected in the normal characterization of percolation. Therefore, it is difficult to determine $\phi_c$ and $\nu$ as defined in Fig.~1 of the main text, in order to perform the scaling collapse. Much larger systems may be required to find out what actually happens for such an attraction.

\begin{thebibliography}{50}
\bibitem{note_range} Although $\mu$ sets both the range and strength of attraction, the range is more relevant than the strength at zero temperature, so we regard $\mu$ as the range of attraction here.
\bibitem{note_2} We notice that some complicated behaviors exist during the evolution from two-scale to one-scale percolation betwen $\mu=10^{-3}$ and $10^{-1}$,
leading to difficulties in a precise characterization of the percolation. For example, for $\mu=10^{-2}$, although we show its $\tau$ vlaue for $f_{\rm p}(\phi, N)=0.5$ in Fig.~\ref{fig:fig3}(d),
the probability of finding rigid states $f_{\rm p}(\phi, N)$ is not well collapsed, as shown in the SM \cite{sm}. More intensive work on much larger systems is required to clear up the evolution from two-scale to one-scale percolation, which is however beyond our current computational capacity.
\end{thebibliography}
\end{document}